\documentclass[prb,twocolumn]{revtex4}
\usepackage{graphicx,amsmath,amssymb,amsxtra}
\renewcommand{\narrowtext}{\begin{multicols}{2} \global\columnwidth20.5pc}
\renewcommand{\v}[1]{{\bf #1}}

\def\be{\begin{eqnarray}}
\def\ee{\end{eqnarray}}
\newcommand{\nn}{\nonumber\\}
\newcommand{\Eq}[1]{Eq.~(\ref{#1})}
\newcommand{\Fig}[1]{Fig.~(\ref{#1})}

\newcommand{\cC}{{\cal C}}
\newcommand{\mq}{{\mathfrak{q}}}
\begin{document}

\title{Domain wall type defects as anyons in phase space}


\author{Alexander Seidel}
\affiliation{National High Magnetic Field Laboratory, Florida State
University, Tallahassee, Fl 32306, USA}
\author{Dung-Hai Lee}
\affiliation{Department of Physics,University of California at
Berkeley, Berkeley, CA 94720, USA}
\affiliation{Materials Sciences Division,
Lawrence Berkeley National Laboratory, Berkeley, CA 94720, USA}

\date{\today}

\begin{abstract}
We discuss how the braiding properties of Laughlin quasi-particles
in quantum Hall states can be understood within a one-dimensional formalism
we proposed earlier. In this formalism the two-dimensional space of the 
Hall liquid is identified with the phase space of a one-dimensional
lattice system, and localized Laughlin quasi-holes 
can be understood as coherent states of lattice solitons. 
The formalism makes comparatively little use of the detailed structure
of Laughlin wavefunctions, and may offer ways to be generalized
to non-abelian states.
\end{abstract}

\maketitle
\section{Introduction}
Fractional quantum Hall liquids are some of the most fascinating
states of matter, displaying topological order, charge
fractionalization, anyon statistics, and non-commutative geometry all
in a real-life laboratory system. Despite these exotic
characteristics, recent research efforts have shown that many of the
fundamental properties of quantum Hall states are adiabatically
rooted in simple one-dimensional charge--density--wave (CDW) states.
This is true for both abelian \cite{seidel1, karl1} and non-abelian
\cite{seidel2,karl2} Hall states. These CDW states appear naturally
when the quantum Hall liquid is studied on a cylinder or torus, and
one circumference of the system is made very small. Although the CDW
states resulting in this limit are trivial and have no dynamics,
they have the same quantum numbers
 as the corresponding fractional quantum Hall states and are
{\em adiabatically connected} to them as the circumference of the
cylinder is increased. This curious aspect of quantum Hall systems
is attractive and useful in a number of ways. On a fundamental
level, it shows that the principle of charge fractionalization in
two-dimensional quantum Hall liquids can be unified with that in
one-dimensional (1d) charge density wave systems.\cite{ssh} On a more
practical level, the correspondence between quantum Hall states and
CDW states offers yet another view of quantum Hall system aside
from the Chern-Simons theories. In addition, it
 reveals a structure of the Hilbert space of
quasi-particle excitations that is not apparent in the traditional
wavefunction approach\cite{laughlin}. It was recently argued by
Haldane\cite{halmarch} that this structure can serve to reduce the
task of obtaining counting rules for hole states considerably.
This has been explicitly demonstrated
 by Read\cite{read} for the case
of clustered nonabelian Hall states.

Recently there is a rising  interest in gaining deeper understanding
of the nonabelian Hall states, fueled by their potential use in
topological quantum computation.\cite{kitaev} The 1d approach
discussed in Refs. \onlinecite{seidel2,karl2} is promising in the sense that
it offers a fresh and simple new way to look at this interesting
state of matter. However, as it is presented so far, this approach
has an important limitation -- it does not offer an obvious way to
understand the braiding statistics of the quasi-particles. The
purpose of this paper is to overcome this limitation for the abelian
quantum Hall liquids. We first resolve the
obvious paradox of how the notion of ``braiding'' can arise in a 1d
formalism. 
The key is to realize that the braiding takes place
in phase-space, which is two-dimensional.
We will then proceed to show how the domain-wall type defects of the
1d formalism acquire the abelian statistics of the anyons in the 
quantum Hall effect.
We will establish this connection in two consecutive steps. The first
approach is physically more intuitive but less rigorous. The second,
more rigorous approach relies heavily on the notion of {\em
duality}, which is an important feature of our 1d formalism for
quantum Hall states.\cite{seidel1,seidel2} We believe that this
route provides a 
pathway that can be used in the case of
non-abelian Hall states as well. 

\section{One-particle quantum mechanics in the lowest Landau level}

We begin by reviewing one-particle quantum mechanics in the lowest
Landau level (LLL) of a torus. We view a torus as a rectangular
strip with dimensions $L_x$ and $L_y$ glued together at opposite
edges. In units where the magnetic length $(Be)^{-1/2}\equiv 1$ (we
shall adopt these units for the rest of the paper), $L=L_xL_y/2\pi$ is
equal to the number of flux quanta passing through the surface of
the torus. In Landau gauge the vector potential is given by \be
\v{A}=(-y,0).\label{landau}\ee In this gauge the vector potential is
single valued as one traverses the torus in the x-direction.
However, it is not single valued in the y-direction. This
non-single-valuedness can be absorbed by a gauge transformation on
the wavefunction so long as the number of flux quanta is integral.
As a result of this gauge transformation, the wavefunction
satisfies periodic boundary conditions in the x-direction, and
the twisted boundary condition $\phi(x,y+L_y)=\exp(-iL_yx)\phi(x,y)$
in the y-direction.\cite{brown} Under the above gauge and boundary
conditions, a complete basis set for the LLL is given by
\be\label{phi1}
  \psi_n(x,y)={\cal N}_\psi\sum_{\ell}\xi^{n+\ell L}
  e^{-\frac 12y^2-\frac 12\kappa^2(n+\ell L)^2}.
\ee Here $\kappa=2\pi/L_x$, $\xi=\exp(-i\kappa z)$, $z=x+iy$, and
${\cal N}_\psi$ is a n-independent normalization constant. The index
$n$ is restricted to range from $0$ to $L-1$, since
$\psi_{n+L}(x,y)=\psi_n(x,y)$. 
\section{The 2D$\leftrightarrow$1D mapping}
The shape of $\psi_n(x,y)$ is that of a ring wrapping around the
$x$-direction of the torus, which is localized to within one magnetic
length around $y=\kappa n$ in the y-direction. By viewing each of
the ring-shaped orbitals as a lattice site, one establishes a mapping
from the Hilbert space of the LLL onto that of a 1d ring of lattice
sites. In this basis, the pseudo-potential Hamiltonians (for which
the Laughlin wavefunctions are exact ground states) become lattice
Hamiltonians describing center-of-mass (CM) conserving pair
hopping\cite{ll,seidel1,karl1} with the hopping range equal to
$1/\kappa$. In the same basis the Laughlin $\nu=1/m$
wavefunctions\cite{laughlin} become lattice wavefunctions of many
particles. In general these lattice wavefunctions are rather complicated.
Great simplification occurs when the limit $L_x\ll 1$ is taken while
keeping $L_y$ large or infinite. In this limit the Laughlin
$\nu=1/m$ wavefunctions describe CDW states where every $m$-th site
is occupied.\cite{hr} In the following, we will represent such a CDW
state by a string of $0$'s and $1$'s, depicting the lattice
occupancy. For example, $1001001001\dotsc$ is a $\nu=1/3$ CDW state.
Clearly there is a 3-fold degeneracy arising from translating this
CDW by one and two lattice constants. What makes the above CDW
states useful is the fact that they are {\it adiabatically}
connected to the quantum Hall liquid states as $L_x$ is
increased.\cite{seidel1,karl1,seidel2,karl2} This adiabatic
evolution allows one to write the $\nu=1/3$ Laughlin state at finite
$L_x$ as \be\label{adia1} |\psi_{1/3}(L_x)\rangle=\,\hat
S(L_x,0)\,|\dotsc1001001001001\dotsc\rangle.\ee Here $\hat
S(L_x,L_x')$ is a unitary operator which transforms a low-energy
state at $L_x'$ into a corresponding state at $L_x$ while keeping
$L_xL_y=2\pi L$ fixed. It is given by \be \,\hat S(L_x,0)\,={\cal
P}_z e^{\int_0^{L_x} dz \hat{D}(z)},\ee where \be
\hat{D}(z)=\sum_{m\ne n} |m(z)\rangle {\langle m(z)|\partial_z
H(z)|n(z)\rangle\over E_n(z)-E_m(z)}\langle n(z)|\label{dz}\ee with
$E_n(z)$ and $|n(z)\rangle$ being the eigen energy and eigenstate of
$H(z)$, the Hamiltonian at $L_x=z$. The operator
 ${\cal P}_z$ is analogous to the time
ordering operator and performs the ``$z$ ordering''. For
definiteness, we shall assume the Hamiltonian in \Eq{dz} to be the
pseudo-potential Hamiltonian used in
Refs. \onlinecite{seidel1,karl1,seidel2,karl2}.

Analogous to \Eq{adia1}, the quasi-particle and quasi-hole
excitations are the adiabatically evolved  domain wall and
anti-domain wall states. For example 
\be 
\hat S(L_x,0)
|..10010010\underline{0}01001001..\!\rangle\label{qh1}
\ee
is a quasihole state. In this way one can establish a one--to--one
correspondence between the low-energy excitations at finite $L_x$
with those at $L_x\rightarrow 0$ analogous to that between the
quasiparticle excitations of a Fermi liquid and the free-electron
excitations of a Fermi gas.

The particular gauge choice in \Eq{landau} breaks the symmetry
between the $x$ and $y$ coordinates of the electron. Had we chosen
the gauge 
\be \v{A}=(0,x)\label{landau2}\ee 
instead, the LLL
orbitals would become 
\be
\psi'_n(x,y)=\psi_n(-y,x)|_{L_x\leftrightarrow L_y}.
\ee
The $\psi'_n$ are ring-shaped
orbitals wrapping around the $y$-direction 
and localized around $\bar\kappa n$ in
the $x$-direction, where $\bar\kappa=2\pi/L_y$.
These orbitals may now be transformed back into the original
gauge \Eq{landau} by means of the following gauge transformation:
\be\label{phi2}
 \bar\psi_n(x,y)&&=\exp(-i xy) \psi'_n(x,y)\nn
&&=
{\cal N}_{\bar\psi}
\sum_{\ell}e^{-\frac 12(x+iy-\bar\kappa(n+L\ell))^2}e^{-\frac 12
y^2} , \ee 
It turns out that this new basis is just the Fourier transform
of the basis \Eq{phi1}. When expressed in the new basis \Eq{phi2},
the pseudo-potential Hamiltonian also describes CM conserving pair hopping, 
except that the
hopping range is now $1/\bar{\kappa}$.\cite{seidel1} In
Refs .\onlinecite{seidel1,seidel2} the transformation relating the two
lattice Hamiltonians is referred as the {\em duality}
transformation.

When the limit $L_y\rightarrow 0$ is taken, 
 the Laughlin wavefunctions again describe
simple CDWs, this time along the $x$-axis of the torus.
For constant $L=L_xL_y/2\pi$ this is equivalent to taking
$L_x\rightarrow\infty$. 
The CDW in this limit can now be expressed in terms
of the orbitals $\bar \psi_n$,  in a manner that is
analogous
to use of the orbitals $\psi_n$ in the opposite limit
discussed above.
Again, the quantum Hall liquid at finite
$L_x$ can be adiabatically evolved
from the CDW at $L_x\rightarrow\infty$, according to  
\be\label{adia2} {|\overline{\psi_{1/3}(L_x)}\rangle}=\,\hat
S(L_x,\infty)\,{|\overline{\dotsc1001001001001\dotsc}\rangle}. \ee
The overbar on the right hand side reminds us that the occupation
numbers are referring to the basis in \Eq{phi2}. The overbar on the
left reminds us that the states in Eqs. (\ref{adia1}) and
(\ref{adia2}) are not identical. This is because the ground state is
three-fold degenerate, and ${|\overline{\psi_{1/3}(L_x)}\rangle}$ in
\Eq{adia2} is a linear combination of the three different ground
states obtained from evolving the three different CDW states in
\Eq{adia1}. Analogous to \Eq{qh1} we can obtain a complete set of 
quasihole states
 as \be\,\hat
S(L_x,\infty)\,{|\overline{..10010010\underline{0}01001001..\!
}\rangle}.\label{qh2}\ee
If one defines the generators of (single particle)
magnetic translations as \be
&&t_x=\exp(-i\bar\kappa p_x)\nn&&t_y=\exp(-i\kappa (p_y+x)),\ee one
can easily check that\be\label{trans}
  t_x\psi_n=&e^{\frac{2\pi i}{L}n}\psi_n,\quad \quad t_y\psi_n&=\psi_{n+1},\nonumber\\
  t_x\bar\psi_n=&\bar\psi_{n+1},\quad\quad \quad t_y\bar\psi_n&
  =e^{-\frac{2\pi i}{L}n}\bar\psi_{n}.\label{trans2}
\ee 
Thus, if the $\psi_n$ are viewed as position eigenstates on a 1d
lattice, the $\bar\psi_n$ are the corresponding momentum eigenstates
and vice versa. This position-momentum duality is a manifestation of
the well known fact that within the lowest Landau level, $x$ and $y$
satisfy a position-momentum type commutation relation ($[x, y]=i$).

We now seek to understand the braiding statistics of Laughlin
quasi-particles in the 1d language established above. For brevity we
consider quasi-holes only. Our main obstacle is that the quasi-hole
states in \Eq{qh1} and \Eq{qh2}  are localized only in one direction
of the torus, and are 
completely delocalized around the other (see below in Section
\ref{cstate}). However the notion of
braiding is only meaningful for point-like quasi-holes that are
localized in both $x$ and $y$.

As $L_x\rightarrow 0$, we can label the $m$ different CDW ground
states by an integer $c=0,1,2,...,m-1$, so that the positions of the
occupied lattice site are given by $n=mp+c$. Here $p=0,...,N-1$ with
$N$ being the particle number. According to \Eq{trans2} these CDW
states are eigenstates of the many-body translation operator
$T_{x}=\prod_i t^i_{x}$, where $i$ is the particle label. A simple
calculation of the eigenvalues gives \be e^{{2\pi i\over
L}\sum_{p=0}^{N-1}(mp+c)}=e^{{i\pi}{(N-1)}}e^{i2\pi{c\over m}}.\ee
In the language of the 1D lattice the above eigenvalue measures the
CM position modulo $L$. Thus for fixed $N$ the center-of-mass
position of the CDW is entirely determined by $c$. Since the
Hamiltonian remains $T_x$-invariant throughout the adiabatic
evolution (changing $L_x$), we conclude \be [\hat
S(L_x,0),T_x]=0.\label{com}\ee \Eq{com} implies that  $c$ continues
to be a good quantum number differentiating the $m$ different ground
states at finite $L_x$.


\section{The Laughlin quasihole as a coherent state of the domain walls\label{cstate}}
\subsection{One quasihole}
The single quasihole states are the ground states of the
system when $L\rightarrow L+1=Nm+1$. In the small $L_x$ limit, such
states can be obtained by inserting an extra empty site into the
ground state CDWs. For example, after inserting an empty site
between the $q$th and $(q+1)$th period of the CM-index-$c$ CDW, the
positions of the occupied sites are given by $n=mp+c$ for $p\le q$,
and $n=mp+c+1$ for $p>q$. The new $T_x$ eigenvalues are given by
\be
&&e^{{2\pi i\over L}\sum_{p=0}^{q}(mp+c)} e^{{2\pi i\over
L}\sum_{p=q+1}^{N-1}(mp+c+1)}\nn&&=e^{{i\pi}{(N-1)}}e^{\frac{2\pi i}{L}
(\frac 1 2(N-1) +Nc-q)}.\label{qhm}
\ee 
The domain wall states characterized by different
$(c,q),~q=0,...,N-1$, are orthogonal. Upon adiabatic evolution each
of them goes into a quasihole state in which the excess charge is
is localized in the $y$ direction, 
but is completely 
delocalized around the $x$ circumference of the torus.

To produce a Laughlin quasihole (localized in both $x$ and $y$
direction) we can turn on a repulsive one-particle potential with
range $\sim$ magnetic length. Such a one-particle potential is
capable of imparting $x$-momentum and hence changing the value
of $T_x$. In view of \Eq{qhm} this clearly
suggests that the Laughlin quasihole state is made of
adiabatically evolved domain wall states with different $q$ but the
same $c$, i.e.,
\be\label{1hole}
 |\psi_c(h)\rangle =\sum_q \,{\varphi^\ast_{c,q}}(h)\, \hat S(L_x,0)
 |c,q\rangle.
\ee In \Eq{1hole} $h=h_x+ih_y$ is the complex coordinate of the hole,
and $|c,q\rangle$ is the domain wall state discussed above.
A physical interpretation for the coefficient $\varphi^\ast_{c,q}(h)$
will be given shortly.

To obtain $\varphi_{c,q}(h)$ we can compute the overlap between
\Eq{1hole} and $|c,q\rangle$. We note that 
\be\langle c,q|\hat
S|c,q'\rangle=0~~~{\rm if~~~}q\ne q' ,
\ee 
since for $q\neq q'$ this is the overlap between two
states of different $T_x$ eigenvalue, according to Eqs.
 (\ref{com}) and (\ref{qhm}). Moreover,
by the $T_y$-translational symmetry, $\langle c,q|\hat S|c,q\rangle$
is independent of $q$. The above arguments imply that
\be\varphi^\ast_{c,q}(h)\propto \langle
c,q|\psi_c(h)\rangle.\label{rw}\ee 
To compute the right hand side of
\Eq{rw} we use the known Laughlin one-quasihole wavefunction
$|\psi_c(h)\rangle$ on a
torus of sufficiently large dimensions. 
In fact, to calculate
this overlap (up to an overall constant), one may work in the simpler 
limit $L_y\rightarrow\infty$
of an infinite cylinder with $L_x$ large but finite.
This is so because the local properties of the Laughlin 
wavefunction $|\psi_c(h)\rangle$ and of the basis \Eq{phi1}, 
which defines the domain wall state $|c,q\rangle$, do
not depend on $L_y$ in the limit where $L_y$ is large.
In this limit the Laughlin wavefunction for a single
quasihole reads
 \be\label{1hole2}
 \psi_{c}(h,\{z_i\})&&=F_c(\{z_i\})\prod_i(\eta-\xi_i)\prod_{i<j}(\xi_i-\xi_j)^m
 e^{-\frac 12\sum_i y_i^2}\nn&&
\ee where $\xi_i=\exp(-i\kappa z_i)$, $\eta=\exp(-i\kappa h)$, and
$F_c(\{z_i\})=(\prod_i\xi_i)^c$.
Since $|c,q\rangle$ is a state where the particle coordinates
$n_i$ take on definite values,
to compute $\langle c,q|\psi_c(h)\rangle$ one only needs to
determine the coefficient of the monomial $\prod_i\xi_i^{n_i}$ in
the polynomial part of
\Eq{1hole2} and 
multiply it by $\exp(\frac 12 \kappa^2\sum_i n_i^2)$ (see
Ref. \onlinecite{hr} for details).
Straightforward calculation gives \be\label{varphi1}
  \varphi_{c,q}(h)={\cal N}^1_{\varphi}\,e^{iq(h_x\kappa+\pi)-
\frac{1}{2m}(h_y-\kappa b_{c,q})^2}\quad, \ee where $ b_{c,q}=m
q+c+{m+1\over 2}.$ 
We note that $\kappa b_{c,q}$ is just the $y$-position
around which the quasi-hole in the state $\hat S(L_x,0)|c,q\rangle$
is localized.
The Gaussian form of \Eq{varphi1} could have been anticipated.
It is analogous to a coherent state in the study of a 1d harmonic
oscillator, which describes a particle that is both localized
in real space as well as in momentum space. 
This is a consequence of the fact that $x$ and
$y$ satisfy a position-momentum type of commutation relation
within the lowest Landau level. It is thus natural that
in our 1d formalism, the $x$ and $y$ coordinates of 
a Laughlin quasi-hole are identified 
with the position and momentum degrees of freedom of our 
1d domain walls.
The position-momentum phase space of these domain walls
can thus be identified with the original 2d configuration
space of Laughlin quasi-particles and quasi-holes
on a torus.

Interestingly, when viewed as a function
of $h$, $\varphi_{c,q}(h)$ can also be viewed as the
wavefunction of a particle of charge $-e/m$, moving under 
the influence of the vector
potential $\v{A}=(-y+\kappa c+\kappa{m+1\over 2},0)$. 
The profile of these orbitals is similar to the
ring-shaped electronic orbitals defined in \Eq{phi1},
except that $\varphi_{c,q}(h)$ is centered around
the domain-wall position $\kappa b_{c,q}$ in the
$y$-direction. These properties appear most natural
if \Eq{1hole} is inverted to express the 
adiabatically continued domain wall states in terms
of localized Laughlin quasi-holes,
\begin{equation}
 \hat S(L_x,0) |c,q\rangle \propto 
\int dh\; \varphi_{c,q}(h)\,|\psi_c(h)\rangle ,
\end{equation}
where $\int dh\,  \varphi_{c,q}(h) \varphi^\ast_{c,q'}(h)\propto\delta_{q,q'}$
was used.


\subsection{Two quasiholes}
A similar strategy can be employed to obtain the expansion of two
localized Laughlin quasiholes in terms of  the adiabatically evolved
two-domain-wall states, i.e.,
 \be
|\psi_c(h_1,h_2)\rangle =&\sum_{q_1<q_2}
\,\varphi^\ast_{c,q_1,q_2}(h_1,h_2)\, \hat S(L_x,0)
|c,q_1,q_2\rangle.\label{2h}\nn&& \ee Here, $|c,q_1,q_2\rangle$
represents the state where two empty sites are inserted into the
CM-index-$c$ CDW. When calculating $\varphi_{c,q_1,q_2}(h_1,h_2)$
some additional thought is necessary. The sum in \Eq{2h} now
contains many terms $|c,q_1,q_2\rangle$ that have the same
$T_x$-eigenvalue, since the latter only depends on $q_1+q_2$. Hence
we cannot use translational symmetry to argue that $\langle
c,q_1,q_2|\hat S(L_x,0)|c,q'_1,q'_2\rangle$ is non-zero only when
$q_{1,2}=q'_{1,2}$.

However, when $\kappa(q_2-q_1)\gg 1$ (i.e., when
the separation between the two domain walls 
is much greater than the
pair hopping range), we expect that the two domain walls
behave as two isolated, non-interacting entities. 
In this case the $q$ value of each domain wall should
remain a good quantum number, and the following matrix element
will factorize:
\be \label{matrixelement}&&\langle c,q_1,q_2|\hat
S(L_x,0)|c,q'_1,q'_2\rangle\nn&&\rightarrow \langle c,q_1|\hat
S(L_x,0)|c,q'_1\rangle \langle c+1,q_2|\hat
S(L_x,0)|c+1,q'_2\rangle\nn&&={\rm
constant}\times\delta_{q_q,q_1'}\delta_{q_2,q_2'}.
\ee 
The shift of
$c$ to $c+1$ in the second factor in the second line 
takes into account that the second domain wall is inserted
into a shifted CDW pattern due to the presence of the first one.
Locally, circumstances are thus as if the second domain wall
had been inserted into the ground state sector $c+1$.
Moreover, for well separated domain
walls
the matrix element \Eq{matrixelement} will
still be diagonal in $q_1$, $q_2$.  
Given this fact we may proceed as in the one-hole
case, and argue that 
\be
\varphi_{c,q_1,q_2}(h_1,h_2)\propto \langle
c,q_1,q_2|\psi_c(h_1,h_2)\rangle.\label{ov}
\ee 
To compute the
overlap on the right hand side of \Eq{ov} we use the Laughlin two
hole wavefunction \be
\psi_c(h_1,h_2,\{z_i\})&=  F_c(\{z_i\})\prod_i(\eta_1-\xi_i)(\eta_2-\xi_i)\nonumber\\
&\times\prod_{i<j}(\xi_i-\xi_j)^m e^{-\frac 12\sum_i
y_i^2}\label{2hole} 
\ee 
A calculation analogous to that for the one
hole case leads to \be\label{varphi2}
  \varphi_{c,q_1,q_2}(h_1,h_2)\simeq{\cal N}^2_{\varphi} \varphi_{c,q_1}(h^-)
\varphi_{c+1,q_2}(h^+) \ee where $\varphi_{c,q}(h)$ is given in
\Eq{varphi1}, and $(h^-,h^+)$ equals $(h_1,h_2)$ for
$h_{1y}<h_{2y}$, and $(h_2,h_1)$ otherwise.
\Eq{varphi2} is valid up to exponentially small
corrections for $\kappa(q_2-q_1)\gg1$.

It is interesting to note that in this limit, the wavefunction
$\varphi_{c,q_1,q_2}$ describes two well separated, non-interacting
particles of charge $-e/m$ in a magnetic field. This is quite analogous
to the one-hole case. However, each
particle sees a slightly different vector potential, namely \be
&&\v{A}_{-}(z)=(-y+\kappa c+\kappa{m+1\over
2},0)\nn&&\v{A}_{+}(z)=(-y+\kappa (c+1)+\kappa{m+1\over
2},0).\label{apm}\ee Thus the second particle feels an additional
flux due to the presence of the first one. This is a manifestation
of the statistical interaction!
This constant shift between the two vector potentials
\Eq{apm} is of a piece with the shift discussed below \Eq{matrixelement}.
Its origin is again the fact that the two domain walls are immersed into
different, mutually shifted local ground state patterns. 
As argued in Refs. \onlinecite{seidel1,karl1,seidel2,karl2}
the shift of the CDW phase due to the presence of a domain wall
is responsible for the fractional charge of the quasi-particles,
by means of the Su-Schrieffer counting argument.\cite{ssh}
We will now show 
that through Eqs. (\ref{apm}), the same shift is also at the
heart of the quasi-particle's fractional abelian braiding statistics.

\section{The Berry phase of exchange: The simple picture\label{exchange1}}

In order to compute the braiding statistics of two quasi-holes we
must calculate the Berry phase along a path taking one hole around
the other while keeping them far separated. A main problem we
encounter in doing so is that for $|h_{1y}-h_{2y}|\lesssim 1$, the
dominant contributions to the right hand side of \Eq{ov} come from
terms with $\kappa(q_2-q_1)\lesssim 1$, where \Eq{varphi2} is not
valid. Such configurations are unavoidable along closed exchange
paths, even though the hole separation may be large at any time.

\begin{figure}
\begin{center}
\includegraphics[width=7cm]{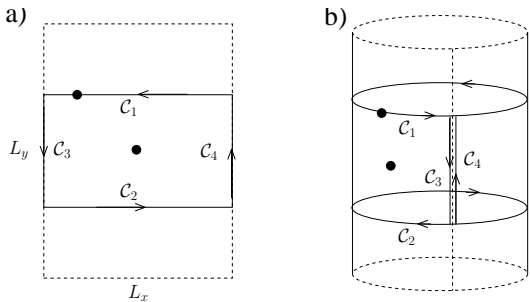}
\caption {The ``special'' braiding path. This path makes use only
of the asymptotic form of the two hole coherent state \Eq{varphi2}.
a) The torus is represented as rectangular strip with periodic
boundary conditions. b) Same as a) with
two opposite ends of the strip in a) glued together.
The two quasi-holes are represented as solid black dots.
}\label{fig0}
\end{center}
\end{figure}

As it turns out, the above difficulty can be overcome in two different ways.
The complete solution to the problem makes use of
the duality discussed earlier, and will be treated in the next 
section. 
However, this approach may obscure
the physical origin of the abelian fractional statistics, which we hinted
above is the simple shift of CDW ground state patterns surrounding
each domain wall. To demonstrate this, we will  first give
a simplified treatment that makes the underlying physics quite transparent.
To do so, we first restrict ourselves to the special class of braiding
paths shown in \Fig{fig0}, where one particle fully encircles
the other. 
These paths contain two pieces 
$\cC_1$ and $\cC_2$ where
one particle fully traverses one hole of the torus, in a regime where
$|h_{1y}-h_{2y}|\gg 1$ holds and \Eq{varphi2} is valid. 
The remaining parts $\cC_3$ and $\cC_4$
of the path cancel in the 
calculation of the Berry phase $\gamma$.
After a simple calculation, 
using Eqs. (\ref{2h}), (\ref{varphi2}),
the latter can then be expressed as:
\be\label{berry0}
\gamma&&=-i\int_{{\cal C}_1+{\cal C}_2}d\v
l\cdot\langle\psi_{c}(h_1,h_2)|\nabla_{h_1}|\psi_{c}(h_1,h_2)\rangle
\nn
&&=\frac{-1}{m}\int_{\cC_1} d \v l \v A_-(h_1)+
\frac{-1}{m}\int_{\cC_2} d \v l \v A_+(h_1)\nn
&&=\frac{-1}{m}\int_{\cC_1+\cC_2} d\v l \v A_-(h_1)
+\frac{-1}{m}\int_{\cC_2} d \v l \v \,\kappa \hat e_x\nn
&&=\frac{-1}{m}\times \,\Phi\,+\frac{2\pi}{m}
\ee 
In here, $\Phi$ is the magnetic flux through the area 
encircled by the closed path in Fig. (\ref{fig0}a),
and we used $\v A_+=\v A_-+\kappa\hat e_x$ in the third line.
\Eq{berry0} is the correct result\cite{asw} for two quasi-holes
in a $\nu=1/m$ liquid. It is now apparent that the abelian statistics
described by the term $2\pi/m$ have their root in the relative displacement
of a ground state CDW pattern in the presence of a domain wall, since
this is the cause for the difference between $\v A_-$ and $\v A_+$.
We can easily generalize this result to the elementary excitations
in hierarchy states. As shown in Ref. \onlinecite{karl1}, the thin
torus limit of a $\nu=p/(mp+1)$ state in the Jain hierarchy\cite{Jain} consists
of unit cells of the form $C=(10_{m-1})_{p-1}10_m$, where a subscript 
denotes repetition of the building block it is attached to.
In the same notation, the expression for a Laughlin state 
such as \Eq{adia1} would be $C=10_{m-1}$.  
Note that for Jain states,
 $m$ is even for fermions, as opposed to our convention for 
Laughlin states. An elementary excitation of charge $-e/\mq$,
where $\mq\equiv mp+1$, corresponds to a domain wall defect in the
thin torus limit where one of the shorter blocks $10_{m-1}$
is removed. In this limit, a state with two domain walls
separated by a number of unit cells would read:
\begin{equation}\label{hierarchy}
 \dotsc \text{CCC} (10_{m-1})_{p-2}10_m \text{CCCCC}(10_{m-1})_{p-2}10_m \text{CCC} \dotsc
\end{equation}
We may now postulate that for a state of two {\em localized} 
charge $-e/\mq$ excitations,
an expression analogous to \Eq{varphi2} holds.
However, two important modifications apply. First,
the definition of $b_{c,q}$ in \Eq{varphi1} should
be properly adjusted to refer to the position
of a domain wall in the Jain state, i.e. 
$b_{c,q}= q\mq +c +\delta$.\cite{note}
Second, it is apparent from \Eq{hierarchy} that the 
shift between the CDW patterns surrounding a domain
wall is now $\Delta c=-m$ rather than $\Delta c=+1$, 
since at each
domain wall one $10_{m-1}$ block of length $m$
is removed. 
Then a calculation analogous to that in
\Eq{berry0}
yields
\be\label{hier1}
  2\theta&&= 2\pi \frac{\Delta c}{\mq}\nn
\ee
for the statistical part of the Berry phase,
with  $\Delta c=-m$ in this case.
 In fact, the arguments leading to \Eq{hier1} can be expected 
to be valid beyond the Jain hierarchy.
That is, whenever
the domain wall picture for an elementary 
charge $-e/\mq$ excitation is known
in an abelian quantum Hall state, 
we expect that its statistical angle $\theta$ satisfies \Eq{hier1}.
Here $\Delta c$ is the shift between CDW patterns
caused by the domain wall, and the filling factor is
$\nu=p/\mq$ with $p$ and $\mq$  coprime. This agrees with the
general result obtained in Ref. \onlinecite{wpsu}.
Indeed, for a charge $-e/\mq$ excitation, the Schrieffer-Su
counting argument implies a relation of the form
\be\label{count}
  \Delta c\, p- n\mq=1\;,
\ee
where $n$ is an integer.\cite{note1}
With this, the result in  Ref. \onlinecite{wpsu} then implies
\be\label{hier2}
   2\theta = 2\pi\frac{\Delta c^2 p}{\mq} \;\mod\;2\pi\; .
\ee
However, using \Eq{count} again, Eqs. (\ref{hier1}) and (\ref{hier2}) agree 
modulo $2\pi$. Note that for a charge $+e/\mq$ defect, \Eq{hier1} requires
an additional minus sign. This corresponds to the
fact that \Eq{varphi1} should be replaced by its complex
conjugate in this case. Since $\Delta c$ will also change sign, the
statistical phase remains the same. 

We should, however, be careful not to apply
\Eq{hier1} to excitations that can be regarded
as composites of more elementary excitations,
such as charge $p/\mq$ Laughlin quasi-particles
in hierarchy states with $p>1$.\cite{note2} For composite particles,
more complicated expressions are generally needed 
to properly localize these particles in both $x$ and $y$.
Once the statistical phases for the most elementary
excitations are known, those of composite particles
can be calculated from the well known composition
rule.\cite{yswu1,yswu2} 
We note, however, that we
have so far determined $\theta$ only modulo $\pi$. 
This is so because we have not carried out a single
exchange of two particles. We will remedy this fact
in the following section. Although we will focus
on Laughlin states from now on for simplicity,
we believe that it is not difficult to generalize
the following discussion to elementary excitations
in arbitrary abelian states.

\section{The Berry phase of exchange: Duality\label{exchange2}}

While the above discussion offers most straight-forward insights
into the nature of abelian statistics from our 1d point of view,
it has certain limitations which we now wish to overcome.
One limitation is the fact that the approach in the preceding
section cannot be generalized to the non-abelian case  even in principle. 
This is because the paths $\cC_3$ and $\cC_4$ in \Fig{fig0} are separated 
by paths $\cC_1$ and $\cC_2$, hence they need not cancel in the
non-abelian case, where all these paths will be represented by non-commuting
matrices.  Also, since braiding statistics are a topological phenomenon,
it is desirable to demonstrate their validity for more general paths, rather
than the special ones considered so far. We will now show how this
is achieved using duality.

The key idea is that in addition to
\Eq{2h}, one could write down a similar expansion for a two-hole
state in terms of adiabatically evolved domain walls from the limit
$L_y\rightarrow 0$ ($L_x\rightarrow\infty$) \be\label{2hole2}
   |\psi_{\bar c}(h_1,h_2)\rangle =\sum_{q_1<q_2} \,\bar\varphi^\ast_{\bar c,q_1,q_2}(h_1,h_2)
   \, \hat S(L_x,\infty)\bigl.\overline{\bigl|{\bar
   c},q_1,q_2\bigr.}\bigr>.\nn&&
\ee  Similar as before, $\bar c$ is the CM index for the CDW in the
$L_x\rightarrow\infty$ limit.  
By going through steps similar to those leading to \Eq{varphi2}, we
can give an asymptotic form for $\bar\varphi_{\bar c,q_1,q_2}$ which
is valid when $\bar\kappa (q_2-q_1)\gg 1$ \be\label{varphi4}
    \bar\varphi_{\bar c,q_1,q_2}(h_1,h_2)\simeq{\cal N}^2_{\bar\varphi}\,
    \bar\varphi_{\bar c,q_1}(h^<)
\bar\varphi_{\bar c+1,q_2}(h^>)  , \ee where \be\label{varphi5}\bar\varphi_{\bar c
,q}(h)=e^{-iq(h_y\bar\kappa+\pi)-\frac{1}{2m}(h_x-\bar\kappa b_{\bar
c,q})^2},\ee
 and  $(h^<,h^>)$ equals $(h_1,h_2)$ for
$h_{1x}<h_{2x}$, and $(h_2,h_1)$ otherwise.  \Eq{varphi4} is valid
for $\bar\kappa(q_2-q_1)\gg 1$, and hence can be used in \Eq{2hole2}
for $|h_{1x}-h_{2x}|\gg 1$.

\begin{figure}
\begin{center}
\includegraphics[width=5.5cm]{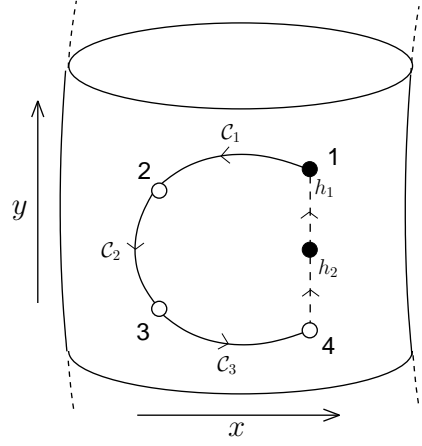}
\caption {Two quasi-holes (solid black) on a torus
are exchanged along a path ${\cal C}$. The position of
one particle is kept at $h_2$, while the other at $h_1$
traces the shown path. The path is divided in to parts
${\cal C}_1$, ${\cal C}_2$, ${\cal C}_3$, separated by
points labeled $1$-$4$.
}\label{fig1}
\end{center}
\end{figure}

In calculating the Berry phase for the exchange of two holes, we can
now employ the following strategy. We start with two holes in the
state $|\psi_c(h_1,h_2)\rangle$, having large $|h_{1y}-h_{2y}|$.
Initially, $h_{1x}=h_{2x}$ is assumed.
Keeping $h_2$ fixed, we move $h_1$ around $h_2$ in a
counter-clockwise manner, dividing the contour ${\cal C}$ into
 three parts (\Fig{fig1}).
Along ${\cal C}_1$, a contribution $\gamma_1$ to the Berry phase can
be calculated using Eqs. (\ref{2h}),(\ref{varphi2}), since
$h_{1y}-h_{2y}\gg 1$ holds. Simple calculation gives \be
\gamma_1&&=-i\int_{{\cal C}_1}d\v
l\cdot\langle\psi_{c}(h_1,h_2)|\nabla_{h_1}|\psi_{c}(h_1,h_2)\rangle
\nn&&=\frac{-1}{m}\int_{{\cal C}_1}d\v l\cdot\v{A}_+(h_1), \ee where
$\v A_+$ is given in \Eq{apm}. At the point labeled 2, we can then
switch to Eqs.(\ref{2hole2}) and (\ref{varphi4}). However before
doing so we need to write $|\psi_c(h_1,h_2)\rangle$ as linear
combinations of $|\psi_{\bar c}(h_1,h_2)\rangle$ as follows
\be\label{P2} |\psi_c(h_1,h_2)\rangle\,\bigr|_2 = \sum_{\bar c}u_{c
\bar c}~~|\psi_{\bar c}(h_1,h_2)\rangle\,\bigr|_2 \ee Along ${\cal
C}_2$, the expressions (\ref{2hole2}),(\ref{varphi4}) can be used to
yield \be \gamma_2(\bar c)&&=-i\int_{{\cal C}_2}d\v
l\cdot\langle\psi_{\bar c}(h_1,h_2)|\nabla_{h_1}|\psi_{\bar
c}(h_1,h_2)\rangle \nn&&=\frac{-1}{m}\int_{{\cal C}_2}d\v
l\cdot\v{B}_-(h_1), \ee where \be &&\v{B}_{-}(z)=(0,x-\bar\kappa
\bar c-\bar\kappa{m+1\over 2})\nn&&\v{B}_{+}(z)=(0,x-\bar\kappa
(\bar c+1)-\bar\kappa{m+1\over 2}).\label{bpm}\ee Since our state
was originally in the sector labeled $c$, we expect that this
remains true even after the adiabatic evolution along ${\cal C}_2$.
At point 3 we must then have \be\label{P3}
  \sum_{\bar c}\bigl. u_{c\bar c}\,e^{i\gamma_2(\bar c)}\left|\psi_{\bar c}(h_1,h_2)\right>\,\bigr|_3
=e^{i\gamma_2}\bigl. \left|\psi_c(h_1,h_2)\right>\,\bigr|_3 .\ee 
We note that 
the validity of
a relation of this form is not trivial,
since  the
constants $u_{c\bar c}$ appearing in it 
are {\em the same} as those defined 
at point 2 in \Eq{P2}.
However, \Eq{P2} is necessarily true if the evolution along
${\cal C}_2$ does not affect the original sector of the
state, which has the label $c$.
For abelian statistics this is what one expects, and we will
verify the validity of \Eq{P3} within our framework below
(see Appendix \ref{app1}, Eqs. (\ref{u1})-(\ref{gamma2}) ).

Finally, for the contribution ${\cal C}_3$ to the Berry phase, we
may again use  Eqs. (\ref{2h}),(\ref{varphi2}). Since now $h_1=h^-$,
the Berry connection is given by $\frac{-1}{m}\v{A}_-(h_1)$, thus
\be \gamma_3&&=-i\int_{{\cal C}_3}d\v
l\cdot\langle\psi_{c}(h_1,h_2)|\nabla_{h_1}|\psi_{c}(h_1,h_2)\rangle
\nn&&=\frac{-1}{m}\int_{{\cal C}_1}d\v l\cdot\v{A}_-(h_1). \ee  As a
last step, we displace both holes vertically until they have exchanged
their original positions, yet this does not contribute to the Berry
phase. Note that, as opposed to the preceding section, our current
path $\cC$ 
describes a true exchange of two particles, not the full encircling
of one particle by the other.
The total Berry phase for this path is given by
\be\gamma=\gamma_1\!+\!\gamma_2\!+\!\gamma_3.\ee By overlapping Eq.
(\ref{P2}) with $|\psi_{\bar c}(h_1,h_2)\rangle|_2$ and
Eq.(\ref{P3}) with $|\psi_{\bar c}(h_1,h_2)\rangle|_3$ we obtain
\be\label{berry2} &&\gamma_2=\gamma_2(\bar c)+\lambda(\bar
c)\nn&&\lambda(\bar c)\!=\! -i\ln\left({\left<\psi_{\bar
c}|\psi_c\right>_2\over\!\left<\psi_{\bar c}|\psi_c\right>_3}\right).
\ee 
Note that by \Eq{P3}, the above result for $\gamma_2$  must  
be independent of the choice for $\bar c$, so long as the
overlaps entering this expression do not vanish. 
Hence the dependence
on $\bar c$ will cancel in the expression for $\gamma_2$.
Writing $\v{B}_-=\v{A}_+
+\nabla f_1$, $\v{A}_-=\v{A}_+ +\nabla f_2$, where
$f_1(z)=xy\!-\!\kappa(c+{m+1\over 2}\!+\!1)x\!-\!\bar\kappa(\bar
c+{m+1\over 2})y$, $f_2(z)=-\kappa x$, the final result for the
Berry phase can be
expressed as \be\label{berry3}
  \gamma&=\gamma_1\!+\!\gamma_2\!+\!\gamma_3=
  -\frac{1}{m}\int_{\cal C}\,d h_1 \,\v{A}_+(h_1)\nonumber\\
&+ [\lambda(\bar c)+\frac{1}{m}
(f_1|_{\substack{\\2}}-f_1|_{\substack{\\3}}+f_2|_{\substack{\\3}}-f_2|_{\substack{\\4}})]
\ee where  $f|_p$ denotes the value of $f$ at point $p$.
The first term in \Eq{berry3} is again equal to \be\label{flux}
&&\frac{-1}{m}\int_{\cal C}\,d h_1
\,\v{A}_+(h_1)=\frac{-1}{m}\times\,\Phi\label{berry35}\ee 
Here $\Phi$ denotes the magnetic flux enclosed by 
the exchange path consisting of the pieces
${\cal C}_1+{\cal C}_2+{\cal C}_3$ and 
the final vertical step
discussed above. This is the Aharonov-Bohm component of the Berry
phase. Our final task is thus to calculate $[\lambda(\bar
c)+\frac{1}{m}
(f_1|_{\substack{\\2}}-f_1|_{\substack{\\3}}+f_2|_{\substack{\\3}}-f_2|_{\substack{\\4}})]$
using \Eq{berry2}. This in turn requires us to compute overlaps of
the form $\left<\psi_{c}|\psi_{\bar c}\right>$ from Eqs.
(\ref{2h}),(\ref{2hole2}).

To achieve this, we need to compute the overlap between $\hat
{S}(L_x,\infty)\,\bigr.\overline{\left|\bar c,p_1,p_2\right.}\bigr>$
and $\hat S(L_x,0) |c,q_1,q_2\rangle$.
The details of the calculation are carried out in Appendix
\ref{app1}. Two cases need to be distinguished. 
In the first case, the particles are either
bosons ($m$ even) of arbitrary particle number $N$, 
or are fermions of odd total
particle number ($m$, $N$ both odd).
The second case is that of
fermions with even particle number
($m$ odd, $N$ even). The latter case is different 
by means of the special role played by the fermion
negative sign in translations, and needs to be treated
separately. This distinction is at the heart
of the ``odd denominator rule'' for fermion quantized
Hall states. How this rule manifests itself
in full generality within our 1d framework
will be discussed elsewhere.\cite{seidel3}
For brevity, we will write down
here the result for the first case (in which
$m(N-1)$ is even):
\begin{equation}\label{overlap}
\begin{split}
&\left<c,q_1,q_2\left|\hat S^\dagger(L_x,0) \hat{S}(L_x,\infty)
\,\right.\overline{\Bigl| \bar c,p_1,
p_2\Bigr.}\right>=\cfrac{\sqrt{m}}{L}\times\\
&
\Big(
\exp\{\frac{i\pi}{m}((2-\frac{2}{L})b_{\bar
c,p_1} b_{c,q_1}+mN(b_{\bar
c,p_1}+b_{c,q_1})+b_{\bar
c\!+\!1,p_2}\\
&+b_{c\!+\!1,q_2}
-\frac{2}{L}b_{\bar c\!+\!1,p_2} b_{c\!+\!1,q_2})\}
+
\exp\{\frac{i\pi}{m}((2\!-\!\frac{2}{L})b_{\bar c,p_1}
b_{c\!+\!1,q_2}\\
&+mN(b_{\bar c,p_1}\! +\!b_{c\!+\!1,q_2})
+b_{c,q_1}
\!-\!b_{\bar c\!+\!1,p_2}\!-\!\frac{2}{L}b_{\bar c\!+\!1,p_2}
b_{c,q_1}\!+\!L)\}
\Big)\,,
\end{split}
\end{equation}
where $b_{c,q}$ is defined below \Eq{varphi1}.
\Eq{overlap} is derived in Appendix \ref{app1}, along with the 
corresponding result for the case $m(N-1)$ odd.
Again, \Eq{overlap} holds only for $\kappa(b_{c,q_2}-b_{c,q_1})\gg
1$, $\bar\kappa(b_{\bar c,p_2}-b_{\bar c,p_1})\gg 1$. These conditions 
are satisfied for the values of $q_1, q_2, p_1, p_2$ that give rise
to the exponentially dominant terms in the overlap
$\left<\psi_{c}(h_1,h_2)|\psi_{\bar c}(h_1,h_2)\right>$ 
at points 2
and 3, because at these points $|h_{1x}-h_{2x}|$ and $|h_{1y}-h_{2y}|$ 
are both large. This overlap can now be calculated by
putting together Eqs. (\ref{2h}), (\ref{varphi2}), (\ref{2hole2})-(\ref{varphi5})
and (\ref{overlap}). In the dominant region of the resulting
sum over $q_i, p_i$
where the Gaussian factors are near their peak, the first
(second) exponential in \Eq{overlap} gives rise to a smoothly
varying phase as a function of the $q_i, p_i$ at point 3 (point 2). 
This term then dominates the sum and can be converted into a Gaussian integral.
This way, one finds the quantities $u_{c\bar c}$
defined via \Eq{P2} at point 2. The explicit expression is given in the Appendix,
\Eq{u1}. It is readily seen that these quantities
form a unitary matrix, as is required. At point 3 the same
procedure shows that a relation of the form \Eq{P3} is
indeed satisfied, involving the same matrix $u_{c\bar c}$ 
obtained at point 2,
and hence the abelian phase $\gamma_2$ is well defined 
(c.f. Appendix \ref{app1}, \Eq{u2}).
The final result for the statistical phase is then:
\be [\lambda(\bar c)+\frac{1}{m}
(f_1|_{\substack{\\2}}-f_1|_{\substack{\\3}}+f_2|_{\substack{\\3}}-f_2|_{\substack{\\4}})]={\pi\over m}\, (\text{mod}\,2\pi)\,.
\label{berry4}\ee 
This, together with Eqs. (\ref{berry3}), (\ref{flux})
is exactly the result expected
for an exchange of two Laughlin quasi-holes. 

\section{Discussion}

In the preceding two sections we have shown how the braiding
properties of Laughlin quasi-holes emerge in a Hilbert space
made up of one-dimensional domain walls. 
The notion of ``braiding'' in these systems arises by
considering coherent states which describe domain walls 
having a narrow distribution both in position space
as well as in momentum space of the 1d system. 
From the 1d point of view, quasi-holes are thus objects 
localized in the phase space of the system.
Since the phase space is two dimensional, the notion of braiding
is well defined, and this phase space is
identified with the original 2d torus of the quantum Hall system.
In this manner, even from a 1d viewpoint one can
recover the well-known fact that
Laughlin quasi-holes behave as fractionally charged anyons in a
background magnetic field, as first derived 
by Arovas, Schrieffer and Wilczek (ASW).\cite{asw}


The fact that it is possible to rephrase the physics
of fractional quantum Hall systems in this 1d language
reflects the 1d structure of Landau levels, and
the related non-commutative geometry of the physics
in a strong magnetic field. However, a brute force
expansion of wave-functions in the Landau level
``lattice'' basis is not very feasible from an
analytic point of view. It is only
due to the remarkable adiabatic continuity between quantum
Hall states and {\em simple} 1d CDW patterns
in the thin torus limit that a simple 1d language
for quantum Hall states can be obtained. 
In earlier works, we had discussed how the intricate quantum
numbers of quantum Hall states may be
inferred from these simple CDW patterns, such as the characteristic
degeneracies and the fractional charge of excitations.
However, the thin torus limit as a starting point seemed
to contain no information about a dynamical principle
that would allow one to understand properties such
as braiding statistics. 
In this work we showed how
the limiting CDW patterns organize the space of 
quasi-hole like excitation such that, when combined
with the duality principle on the torus,
the statistics of Laughlin-quasi holes can be 
derived. While the underlying calculation is less 
straightforward than the standard ASW procedure,
it has a number of attractive features. Unlike
in the ASW treatment, it is quite apparent
in our method that corrections to the Berry phase
\Eq{berry4} will die off exponentially with the radius
of the exchange path.
In the traditional approach 
this fact had been obscured by the finite extent of 
the quasi holes.\cite{asw} 
More importantly, the ASW treatment makes use 
of the very special product structure of Laughlin-type wavefunctions.
Therefore it cannot be carried over to more complicated cases in principle.
While we have not yet shown that our method makes the
calculation of non-abelian statistic a workable task,
we believe that the general structure of our approach
does carry over to the non-abelian case.
To be specific, we point out that the local Berry connections 
which arise in the basis of adiabatically continued
domain-wall states are locally trivial,  describing
a charge $-e/m$ particle interacting with a constant
magnetic background field (Eqs. (\ref{apm}), (\ref{bpm})).
We conjecture that this feature also holds in the
non-abelian case. 
That is, we conjecture that in the basis of adiabatically
continued domain wall states, the local Berry connection
will be diagonal along each path segment in \Fig{fig1}.
The information about
non-abelian statistics would then be entirely contained
in the transition functions describing the 
change between the two mutually dual sets of basis 
wavefunctions at points 2 and 3, which generalize
Eqs. (\ref{P2}) and (\ref{P3}) of this paper to the
non-abelian case. 
This may considerably reduce the task of 
deriving non-abelian statistics directly from
wavefunctions. 
We reserve a detailed
study of this conjecture for future work.

\section{Conclusion}
To conclude we have, in previous works, established that the
fractional charge of the abelian and non-abelian quasiparticles can
be understood as the fractional charge carried by the solitons in
appropriate one dimensional CDW systems (Refs. \onlinecite{seidel1, seidel2},
c.f. also Refs. \onlinecite{karl1,karl2}). In this paper we show that the fractional statistics of the
abelian quasiparticles can also be understood in this language. The
bottom line is that the Laughlin quasiparticle state is the {\it
coherent state} of the one-dimensional solitons. We have explicitly
calculated the expansion coefficient of such coherent states in terms
of the position eigenstates of the solitons. 
These expansions were obtained by making contact with Laughlin
wavefunctions. However, their simple structure seems to
follow naturally from the symmetries of the problem, as
well as the non-commutative geometry of the physics in a
strong magnetic field. The latter leads to the identification
of the two-dimensional surface of a quantum Hall torus with
the position-momentum phase space of a one-dimensional (lattice)
system. The role of localized Laughlin quasi-particles is then
naturally identified with that of coherent states formed by
domain-wall type excitations in the 1d picture. 
Based on this approach we were able to give a new derivation
for the statistics of Laughlin quasi-particles. This approach
also makes use of the inherent duality of the 1d formalism,
and does not appear to be as closely tied to the specific
structure of Laughlin wavefunctions when 
compared to the traditional approach.
We are hopeful that the formalism presented here 
will offer ways to calculate the braiding statistics
for non-abelian states, where the traditional many-body
wavefunctions are far more complicated.\cite{mooreread, readrezayi}

\acknowledgements

One of us (DHL) was supported by
  the U.S. Department of Energy Grant
  DE-AC03-76SF00098.

\appendix
\section{\label{app1}Relation between dual representations and other details}

In this Appendix we derive 
the relation between
states that are obtained by the adiabatic evolution
of two-domain-wall states from opposite thin torus limits, i. e.
$L_x\rightarrow 0$ and $L_y\rightarrow 0$, respectively.
This relation is needed to calculate the overlap in \Eq{overlap}.
We first solve the analogous problem
for single hole domain wall states on a torus with $L=mN+1$ flux quanta.
Thus we seek an expression of the form
\be \label{expan0}
\hat{S}(L_x,\infty)\overline{|\bar c, p}\rangle=\sum_{c,q} u(\bar c,p|c,q)
\hat {S}(L_x,0)\bigl|{c,q}\bigr>
\ee 
Note that the  sum on the right hand side is {\em not} restricted
to a single $c$-sector. Hence to ease the notation and the expressions
that follow, we will now label states by the domain-wall positions
\be\label{ab}
   b\equiv b_{c,q}=mq+c+\frac 12(m+1)
\ee
and
\be
   a\equiv b_{\bar c,p}=mp+\bar c+\frac 12(m+1)
\ee
respectively. Recall that $\bar\kappa a$ and $\kappa b$ are
just the $x-$ and $y-$positions of a domain-wall in the states
\be
  \overline{|a}\rangle\equiv\overline{|\bar c,p}\rangle
\;,\; |b\rangle\equiv |c,q\rangle
\ee  
respectively. We shall also use the abbreviations
\be\label{S}
\hat S\equiv \hat S(L_x,0)\;,\;\;\hat{\bar S}\equiv \hat S(L_x,\infty)
\ee
in the following.
With the notation 
Eqs. (\ref{ab}), (\ref{S}) we can rewrite \Eq{expan0} in
the more compact form
\be \label{expan2}
\hat{\bar S}\overline{|a}\rangle=\sum_{b} u(a|b) 
\,\hat {S}\bigl|{b}\bigr>\; .
\ee 
Note that from \Eq{ab}, $a$ and $b$ are both integer for
$m$ odd, and half-odd integer for $m$ even. The sum in
\Eq{expan2} thus goes over all integer {\em or} half-odd integer
values in the interval $[0,L)$, depending on $m$.

We now observe that
$\hat{\bar S}{\bigl.\overline{\bigl|a\bigr.}\bigr>}$
is an eigenstate of $T_y$ with eigenvalue
$\exp(-ik)$, $k=-\frac{2\pi}{Lm}a +\pi(N+\frac{2}{m}a)$.
This implies that $\hat{\bar S}{\bigl.\overline{\bigl|a\bigr.}\bigr>}$
has the form of a plane wave in terms of the states
$\hat {S}\bigl|{b}\bigr>$,i.e.
\be\label{a1}
  \hat {\bar S}{\bigl.\overline{\bigl|a\bigr.}\bigr>}
\,\propto\frac{1}{\sqrt{L}}\sum_b  e^{ikb}\hat {S}\bigl|{b}\bigr> .
\ee
For the time being, we restrict ourselves to
the cases where the underlying particles are either bosons
($m$ even, $N$ arbitrary), or fermions with an odd total number of particles
($m$, $N$ both odd). In both these cases the domain wall
states transform straightforwardly under translations,
\be\label{trans3}
T_y |{b}\bigr>=\left\{
\begin{array}{ll}
|{b+1}\bigr> & \text{for $b+1<L$}\\
|{b+1-L}\bigr> & \text{otherwise}
\end{array}\right.
\ee
The situation is slightly more complicated for fermions when
the particle number $N$ is even.
This is so because then an additional
minus sign arises whenever an electron in the state $|b\rangle$
is translated across the boundary between the orbital labeled $L-1$ 
and the orbital labeled $0$. Said succinctly, we distinguish the following
two cases:
\begin{equation}\label{cases}
\begin{split}
\text{case~~i)}&\quad m(N-1)\;\; \text{even}\\
\text{case ii)}&\quad m(N-1)\;\; \text{odd}
\end{split}
\end{equation}
where we leave it understood that the underlying
particles are fermions if $m$ is even, and bosons
if $m$ is odd. We first consider case i) 
where \Eq{trans3} holds, and deal with case ii) later.
From \Eq{trans3} it is easily
verified that the right hand side of \Eq{a1} has the correct 
$T_y$ eigenvalue, 
since $T_y$ commutes with the evolution operator
$\hat S$.
The expression \Eq{a1} is, however, not complete yet. We must still
choose the overall phase of the right hand side in a consistent manner.
The correct phase as a function of $a$ can be determined from the
requirement that
$T_x \hat{\bar S}{\bigl.\overline{\bigl|a\bigr.}\bigr>}=
\hat{\bar S}{\bigl.\overline{\bigl|a\!+\!1\bigr.}\bigr>}$. Alternatively,
using duality it can be shown that
$u(a|b)$
must be symmetric in $a$ and $b$, i.e. $u(a|b)=u(b|a)$. Both requirements
yield that the overall phase factor in \Eq{a1} must be $\exp(i\pi Na)$.
Altogether, this results in
\be\label{overlap2}
&&{\bigl<b\bigr|}\hat S^\dagger
 \hat{\bar S}{\bigl.\overline{\bigl|a\bigr.}\bigr>}
=u(a|b) \nn
&&=\frac{1}{\sqrt{L}}\exp\{\frac{i\pi}{m}((2-\frac{2}{L})ab+mN(a+b))\}
\ee 

We now turn to the actual two-hole problem on a torus with
$L=mN+2$ flux quanta. Let us seek an expansion for $\hat{\bar
S}{\bigl.\overline{\bigl|a_{1}, a_{2}\bigr.}\bigr>}\equiv \hat{\bar
S}{\bigl.\overline{\bigl|\bar c,p_1,p_2\bigr.}\bigr>}$ 
in terms of the states
$\hat{S}{\bigl.{\bigl|b_{1}, b_{2}\bigr.}\bigr>}\equiv\hat{S}{\bigl.{\bigl|c,q_1,q_2\bigr.}\bigr>} $,
\be\label{u}
\hat{\bar S}{\bigl.\overline{\bigl|a_1, a_2\bigr.}\bigr>}
=\sideset{}{'}\sum_{b_1<b_2} u(a_1,a_2|b_1,b_2)
\hat{S}{\bigl.{\bigl|b_1, b_2\bigr.}\bigr>} .
\ee
Again, $m$ determines whether the sum goes over integer
or half-odd integer values. In addition, the following
constraints apply:
\be
0\leq a_1<a_2<L\,,&&\;0\leq b_1<b_2<L\nn
a_2-a_1\equiv 1 \mod m\,,&&\;b_2-b_1\equiv 1 \mod m\label{constraint}
\ee
The second line expresses the fact that the second domain-wall
is inserted into a CDW pattern which is shifted by one lattice
site relative to the pattern surrounding the first domain wall.
The prime on the sum in \Eq{u} denotes that the 
constraint \Eq{constraint} is enforced.


As in deriving \Eq{varphi2}, we are
facing the problem that the matrix elements 
$u(a_1,a_2|b_1,b_2)$
are not entirely determined by translational symmetry alone.
To make progress, we first of all assume that the domain wall
positions $a_1$ and $a_2$ are well separated, i.e. 
$\bar\kappa(a_2-a_1)\gg 1$, such that the ``dressing''
of each domain wall by the operator $\hat{\bar S}$
will be unaffected by the presence of the other domain wall.
The two defects are then independent. 
When the separation of the  
$\kappa(b_2-b_1)$ 
in the expansion \Eq{u} is also large, we expect
that the expression
in \Eq{u} should be of a plane-wave form
analogous to \Eq{a1} in both variables $b_1$ and $b_2$.
We thus write down an ansatz of the form
\be\label{u2}
  u(a_1,a_2|b_1,b_2)\simeq {\cal N}_u e^{\beta(a_1,a_2)}
\!\!\left(e^{ik_1b_1+ik_2 b_2}\!+\!e^{i\lambda}e^{ik_1b_2+ik_2b_1} \right)
\nn
\ee
for $\kappa(b_2-b_1)\gg 1$, $\bar\kappa(a_2-a_1)\gg 1$, where we must
now determine the parameters $\beta$, $k_1$, $k_2$, $\lambda$ as a
function of $a_1$, $a_2$, $b_1$, $b_2$.
We first use translational symmetry. One finds that 
$\hat{\bar S}{\bigl.\overline{\bigl|a_1, a_2\bigr.}\bigr>}$
is an eigenstate of $T_y$ with eigenvalue 
$\exp(-iK)$, where
\begin{equation}\label{K}
  K=\frac{\pi}{m}(-\frac{2}{L}(a_1+a_2)+Nm+2a_1+\eta) \mod 2\pi \,.\,
\end{equation}
In the above, the constant $\eta\equiv 1$ comes from the constraint
$a_2-a_1\equiv 1\mod m$ in \Eq{constraint}. 
It is useful to introduce this dummy variable,
since one would  naively expect that
$a_1$ and $a_2$ should enter expressions such as \Eq{K} more 
symmetrically.
Due to the form of the constraint however, the expression is
truly symmetric only under the exchange $a_1\leftrightarrow a_2$
and the simultaneous substitution $\eta\rightarrow -\eta$.
Although this symmetry is not immediately
obvious in equation \Eq{K}, it is easily checked that it is
satisfied (modulo $2\pi$). 
Similar statements hold for some of the expressions that
will follow, hence $\eta$ is best retained as a variable for easy
consistency checks. Furthermore,
we also note that $\exp(iKL)=1$ holds
as required by periodic boundary conditions.
The requirement that the right hand side of \Eq{u} must also be
a $T_y$ eigenket with eigenvalue $\exp(-iK)$ leads to the
following conditions:
\begin{align}
  &k_1+k_2=K \mod 2\pi\label{ksum}\\
  &\lambda=-k_1L=k_2L \mod 2\pi\label{lambda}
\end{align}
where in the last line, it was used that
$T_y \hat{S}{\bigl.{\bigl|b_1, b_2\bigr.}\bigr>}=
\hat{S}{\bigl.{\bigl|b_2\!+\!1\!-\!L, b_1\!+\!1\bigr.}\bigr>}$ 
holds when $b_2+1>L$.
We now determine how the phase factors $e^{ib_jk_{j'}}$ in \Eq{u2}
must change when a domain wall undergoes a local move.
For this we first need to make precise
what a local move is. 
We stress again that it is not possible for any 
domain wall to change its position $b_i$ by an amount
smaller than $\pm m$ without shifting the entire
fluid, i.e. without changing the low energy sector $c$.
This is evident from Eqs. (\ref{ab}). Thus it is clear
that a change of any domain-wall position by an amount
$\Delta b<m$ is not a ``local'' move, but affects an 
infinite number of degrees of freedom.
In contrast, a domain-wall move by $\Delta b=\pm m$ only requires 
the hopping of a single electron in the thin torus limit.
Even for the ``dressed'' domain walls at finite circumference,
we expect that a local operator (such as the local charge density
operator) will be able to generate matrix elements only
between states 
$\hat{S}{\bigl.{\bigl|b_1, b_2\bigr.}\bigr>}$
whose domain wall positions $b_1$ or $b_2$ differ by a few {\em integer}
multiples of $m$. Hence it is the change of  the phase 
factors $e^{ib_jk_{j'}}$ in \Eq{u2} under a change of $b_j$ by
$\pm m$ that will determine physical properties like the charge density
profile of the state \Eq{u}. Let us consider the single hole case, \Eq{a1}.
We note that in a state describing a hole localized at
$h_x\equiv\bar\kappa a$, the phase of $\hat {S}\bigl|{b}\bigr>$
always changes by the following amount 
when $b\rightarrow b+m$:
\be\label{phase}
 e^{ikm}&&=e^{-2\pi i a/L+i\pi(Nm+2a)}\nn
&&=-e^{-2\pi i a/L}
=-e^{-i\kappa h_x},
\ee
where we have used that $Nm+2a$ is always odd in case i) (\Eq{cases})
which we are considering here.
Incidentally, the same change of phase under $b\rightarrow b+m$ as
that shown in \Eq{phase} can already be observed in
the single hole coherent state Eqs. (\ref{1hole}),(\ref{varphi1}),
and more importantly so in the two-hole coherent state
Eqs. (\ref{2hole}),(\ref{varphi2}). 
It is thus quite clear that 
we must also have
\be\label{kj}
  e^{ik_{1,2}m}=-e^{-2\pi i a_{1,2}/L}
\ee
in \Eq{u2}, in order for the state \Eq{u} to describe
two dressed domain walls at $x$-positions $\bar\kappa a_{1,2}$. 
The conditions Eqs. (\ref{K}),(\ref{kj}) are satisfied
by the following choice of 
the momenta $k_1$, $k_2$,
\begin{equation}
\begin{split}\label{k1k2}
k_1&=\frac{\pi}{m}(-\frac{2}{L}a_1+2a_1+Nm)\\
k_2&=\frac{\pi}{m}(-\frac{2}{L}a_2+\eta)\;.
\end{split}
\end{equation}
Superficially, it looks like one could have made different choices
for $k_1$, $k_2$ that also satisfy Eqs. (\ref{K}),(\ref{kj}).
However, using the constraint \Eq{constraint}
it can be shown that all these choices give rise to the same state,
up to a trivial overall phase. In general, one may let
$k_1\rightarrow k_1\!+\!\Delta$, $k_2\rightarrow k_2\!-\!\Delta$, 
where $\Delta$ is an integer multiple of $2\pi/m$, without
changing the state \Eq{u}. 
In particular, the state \Eq{u}
is invariant (up to a phase) when all
indices $1$ and $2$ are exchanged in \Eq{k1k2}, 
and the
substitution
$\eta\rightarrow -\eta$ is made.
Finally, we fix the overall phase of the state by choosing $\beta(a_1,a_2)$
in \Eq{u2}. Again we do this by requiring that the state \Eq{u}
transforms properly under $T_y$ translations, i.e.
$T_y\hat{\bar S}{\bigl.\overline{\bigl|a_1, a_2\bigr.}\bigr>}
=\hat{\bar S}{\bigl.\overline{\bigl|a_1+1, a_2+1\bigr.}\bigr>}$,
and that the matrix element $u(a_1,a_2|b_1,b_2)$ is symmetric
under the simultaneous exchange
$a_1\leftrightarrow b_1$, $a_2\leftrightarrow b_2$,
as required by duality. This way one obtains
\be\label{beta}
 \beta(a_1,a_2)=\frac{\pi}{m}( Nm\, a_1+\eta\, a_2)  .
\ee
With this choice, the first term in \Eq{u2}
is manifestly symmetric under the exchange
$a_j\leftrightarrow b_j$, and the second term
can be shown to have this symmetry using 
again \Eq{constraint}.
Plugging Eqs. (\ref{beta}),(\ref{k1k2}), (\ref{lambda}),
and $\eta\equiv 1$
into \Eq{u2} yields the matrix element
$u(a_1,a_2|b_1,b_2)$ displayed in \Eq{overlap}.
In writing \Eq{overlap}, 
we also used that due to the asymptotic 
plane wave form of $u(a_1,a_2|b_1,b_2)$, 
the normalization ${\cal N}_u$ 
must be equal to the square root of the number of terms
in \Eq{u2}, at least to the leading order in $1/L$.
This yields ${\cal N}_u \simeq \sqrt{m}/L$. 
Although this result
does not enter our determination of the Berry phase,
it is interesting to note that 
corrections to it are actually 
exponentially small. This can be shown
from the requirement that the matrix formed by the quantities
$u_{c \bar c}$ in \Eq{P2} must be unitary, as we 
will see shortly.
We had refrained from giving a detailed expression
for these quantities in the main text for brevity.
This expression will be given in the following.
By carrying out the procedure described in Section
\ref{exchange2} at point 2 (\Fig{fig1}), one obtains:
\be
 u_{c \bar c}&&= \frac{L^2}{2m^2}{{\cal N}^1_{\varphi}{{\cal N} ^2_{\varphi}}
{{\cal N}_u}\exp\{-\frac{i}{m}[\pi mN(\bar c+c+m+2)}\nn
&&+\pi m(\bar c+c+m)
+\pi m+ 2\pi(c+\frac{m+1}{2})(\bar c+\frac{m+1}{2})\nn
&&-\kappa(c+\frac{m+1}{2})(h_{1x}+h_{2x})
-\bar\kappa(\bar c+\frac{m+1}{2})(h_{1y}+h_{2y})\nn
&&-\kappa h_{1x}-\bar \kappa h_{2y} +h_{1x}h_{1y}+h_{2x}h_{2y}]\}\nn
&&\equiv\langle \psi_{\bar c}(h_1,h_2) |\psi_c(h_1,h_2)\rangle\bigr|_2
\label{u1}
\ee
It is easily seen that the above expression is proportional
to a unitary matrix. Since we are interested in domain
walls that are well separated in $x$ and $y$, one can 
determine the normalization constants ${\cal N}^1_{\varphi}=\kappa\sqrt{m/\pi}$
and  ${\cal N}^2_{\varphi}=\bar\kappa\sqrt{m/\pi}$ from 
Eqs.  (\ref{varphi2}) and (\ref{varphi4}) up to exponentially small 
corrections. The condition that $u_{c \bar c}$ is unitary then yields 
${\cal N}_u=\sqrt{m}/L$, as anticipated. 
At point 3 one may calculate the overlap between the states
$\psi_c$ and $\psi_{\bar c}$ using the same procedure.
This defines the phase $\gamma_2$, as explained in Section 
\ref{exchange2}. One finds:

\be
\langle \psi_{\bar c}(h'_1,h_2) |\psi_c(h'_1,h_2)\rangle\bigr|_3
=u_{c\bar c}|_{\substack{\\h_1\rightarrow h'_1}}\nn
\times \exp\{-\frac{i}{m}[\pi+\kappa (h'_{1x}-h_{2x})]\}
\label{u3}
\ee
where $h_1$ and $h'_1$ are the positions of the moving
particle at point 2 and point 3, respectively. In the above, all occurrences
of $h_1$ in $u_{c\bar c}$ are replaced by $h'_1$. However,
this result can be recast to be of the form \Eq{P3}, with the
{\em original} $u_{c\bar c}$ defined at point 2, and with
\begin{equation}\label{gamma2}
 \gamma_2=\frac{-1}{m}\int_{{\cal C}_2}d\v l\cdot\v{A}_+
+\frac{1}{m}(\kappa h'_{1x}-\kappa h_{2x}+\pi) . 
\end{equation}
When this result is plugged into \Eq{berry3}
 the final result \Eq{berry4} is obtained.

Finally, we comment on how the above equations 
need to be modified in case ii), 
which corresponds to an even number
of fermions ($m$ odd, $N$ even).
In this case a single domain-wall state $|b\rangle$
represents a Slater determinant state that does
not quite obey the simple transformation law
\Eq{trans3} under translations. Rather, an additional
negative sign must be introduced on the right hand side
whenever an electron in the state $|b\rangle$ is translated
from site $L-1$ to site $0$. 
This is so because an
electron creation operator has to be commuted through
$N-1$ other such operators in this case. 
This happens 
every $m$ translations, except when the domain wall
itself moves across the boundary. 
The
properties of these states under translation are thus 
slightly more complicated. For single domain-wall states,
however, the additional phase can be removed simply 
by considering multiplication with the following prefactor:
\be \label{prefactor}
   e^{i\pi q} |b\rangle\;.
\ee
In here, $q\equiv q_b$ is the integer related to $b$ via \Eq{ab}.
Note that $q_b$ is uniquely defined by the requirement
$c \in \{0,...,m-1\}$. It is easily checked that the
prefactor compensates for the fermion minus sign in 
translations, and the states \Eq{prefactor} transform
under $T_y$ translations in a manner analogous to
\Eq{trans3},
\be\label{trans4}
T_y  e^{i\pi q_b}|{b}\bigr>=\left\{
\begin{array}{ll}
 e^{i\pi q_{b+1}}|{b+1}\bigr> & \text{for $b+1<L$}\\
 e^{i\pi q_{b+1-L}}\,|{b+1-L}\bigr> & \text{otherwise,}
\end{array}\right .
\ee
where the last line uses the fact that $N=(L-1)/m$ is even.
It follows that with the following modification of the
amplitude $u(ab)$ in Eqs. (\ref{expan2}), (\ref{overlap2}),
\be
 u(a|b) \rightarrow e^{i\pi(q_a+q_b)}\,u(a|b)\;,
\ee
the right hand side of \Eq{expan2} still has the
 correct properties under the action of
$T_x$ and $T_y$, which uniquely define the states
$\hat{\bar S}\overline{|a}\rangle$.
We observe that thanks to the additional factor,
the change of phase under local moves is still
the same as one expects from \Eq{phase}, i.e
\be\label{move}
 u(a|b+m)=-e^{-2\pi ia/L} u(a|b)\;.
\ee
For two domain-wall states, we can now construct a 
$T_y$ eigenstate by modifying Eqs. (\ref{u}) and
(\ref{u2}) as follows:
\be\label{u4}
&&u(a_1,a_2|b_1,b_2)\simeq\nn
&&{\cal N}_u e^{\beta'(a_1,a_2)}
\!\!\left(e^{ik_1b_1+ik_2 b_2+i\pi q_{b_1}}\!+\!e^{i\lambda}e^{ik_1b_2+ik_2b_1+i\pi q_{\{b_2\!-\!1\}}} \right)\nn
&&\equiv
u_1(a_1,a_2|b_1,b_2)+u_2(a_1,a_2|b_1,b_2)\;,
\ee
where $u_1$ and $u_2$ denote the first and
second term in the second line, respectively.
With this choice, the state \Eq{u} is still a
 $T_y$ eigenstate of eigenvalue $\exp(-iK)$,
where the parameters $k_1$, $k_2$ and $\lambda$
are still subject to the conditions Eqs.
(\ref{ksum}), (\ref{lambda}). Again,
$k_1$, $k_2$ are determined from \Eq{K}
and the analogue of \Eq{kj}, which is the
condition that the two terms in
\Eq{u4} behave analogous to the phases of the 
single hole case under local moves, as determined
in \Eq{move}:
\begin{equation}
\begin{split}
u_{1,2}(a,_1,a_2|b_1+m,b_2)&=-e^{-2\pi ia_{1,2}/L} u_{1,2}(a,_1,a_2|b_1,b_2) \\
u_{1,2}(a,_1,a_2|b_1,b_2+m)&=-e^{-2\pi ia_{2,1}/L} u_{1,2}(a,_1,a_2|b_1,b_2)\;.
\end{split}
\end{equation}
It turns out that these conditions are satisfied by the same
choices for $k_i$ made above in \Eq{k1k2} 
(where the term $\pi N$ may be dropped). 
The necessary adjustment
to the overall phase $\beta'(a_1, a_2)$ is:
\be\label{beta2}
 \beta'(a_1,a_2)=\frac{\pi}{m}(a_1+\eta\, a_2)+\pi q_{a_1}  .
\ee  
With this, 
the matrix element $u(a_1,a_2|b_1,b_2)$
and the resulting 
states $\hat{\bar S}\overline{|a_1,a_2}\rangle$
 have exactly the same symmetries 
and translational properties discussed above for case i).
We have verified that with these modifications, the Berry phase
calculation along the lines discussed above again yields the correct
result \Eq{berry4} in case ii).


\begin{thebibliography}{99}
\bibitem{note}
A suitable choice for $\delta$ would be
$\delta=m(p-2)/2$, referring to the center of
the truncated $(10_{m-1})_{p-2}1$ block.
However, the precise value of $\delta$ is not
crucial. Likewise, the coefficient $1/2m$ in
the exponent of \Eq{varphi1} may be adjusted.
However, this coefficient only affects the shape
of the quasi-hole. Note that the overall $1/m$
prefactor in \Eq{berry0} comes from $(\partial b_{c,q}/\partial q)^{-1}$,
which is now changed to $1/\mq$.
\bibitem{note1}
This can be seen as follows:
Consider a charge density wave pattern with no defects
on a lattice with $L$ sites and $N$ particles, such that
$\nu=N/L=p/\mq$. Then insert a charge $-e/\mq$ domain wall,
changing the system size to $L+\Delta c$ and the number
of particles to $N+n$. The charge balance then reads
$N+n=\nu(L+\Delta c)-1/\mq$. This is equivalent to \Eq{count}.
\bibitem{note2}
The domain wall picture for a Laughlin quasi-hole
is generally given by inserting an additional zero anywhere
into the ground state CDW pattern.
\end{thebibliography}
\end{document}